\documentclass[superscriptaddress, amsmath, amssymb, prx, nofootinbib, twocolumn
]{revtex4-2}
\usepackage{graphicx}
\usepackage{comment}
\usepackage{dsfont}
\usepackage{color}
\usepackage{bm} 
\usepackage{hyperref}

\usepackage[utf8]{inputenc}
\usepackage{amsmath}
\usepackage{amssymb}
\usepackage[a4paper, total={6.6in, 8.8in}]{geometry}
\usepackage{float}
\usepackage{braket}
\usepackage{adjustbox}
\usepackage{color}
\usepackage{framed} 
\usepackage{url}
\usepackage[normalem]{ulem}
\usepackage{subfigure}

\begin{document}

\begin{abstract}
We present a method to approximately solve general instances of combinatorial optimization problems using the physical dynamics of 3d rotors obeying Landau-Lifshitz-Gilbert dynamics. Conventional techniques to solve discrete optimization problems that use simple continuous relaxation of the objective function followed by gradient descent minimization are inherently unable to avoid local optima, thus producing poor-quality solutions. Our method considers the physical dynamics of macrospins capable of escaping from local minima, thus facilitating the discovery of high-quality, nearly optimal solutions, as supported by extensive numerical simulations on a prototypical quadratic unconstrained binary optimization (QUBO) problem. Our method produces solutions that compare favorably with those obtained using state-of-the-art minimization algorithms (such as simulated annealing) while offering the advantage of being physically realizable by means of arrays of stochastic magnetic tunnel junction devices.
\end{abstract}

\title{Solving combinatorial optimization problems through stochastic 
Landau-Lifshitz-Gilbert dynamical systems}
\author{Dairong Chen}
\affiliation{Center for Quantum Phenomena, Department of Physics, New York University, New York, NY 10003 USA}
\author{Andrew D. Kent}
\affiliation{Center for Quantum Phenomena, Department of Physics, New York University, New York, NY 10003 USA}
\author{Dries Sels}
\affiliation{Center for Quantum Phenomena, Department of Physics, New York University, New York, NY 10003 USA}
\affiliation{Center for Computational Quantum Physics, Flatiron Institute, New York, NY 10010 USA}
\author{Flaviano Morone}
\affiliation{Center for Quantum Phenomena, Department of Physics, New York University, New York, NY 10003 USA}

\maketitle

\emph{Introduction.}
A compelling idea linking physics and discrete 
mathematics is to leverage the intrinsic dynamics 
of real-world physical systems to solve mathematical 
optimization problem. The idea relies on correspondence 
between the variables of the mathematical 
problem and the degrees of freedom of an analogue physical 
system, where the role of the objective function 
to be optimized is played by the energy function 
of the interacting physical system~\cite{mezardmontanari, krzakala, moore, Fu_1986}. 
Once this correspondence has been successfully 
established, all one has to do is to cool 
the system down to zero temperature and wait 
for it to settle in its ground state, 
{\em i.e.}, the state with minimal energy, which, 
in turn, represents the solution to the original 
optimization problem~\cite{vecchi}. 
In many cases of practical interest, this is easier 
said than done in that during the cooling 
process the physical system can get stuck in 
long-lived metastable states, possibly without 
ever reaching the ground state on human-life 
timescales. 
In such situations, the choice of the analogue physical 
system becomes the crucial point of the entire methodology. 
Physical systems that are less prone to 
getting trapped in the local minima of the 
energy function will solve the optimization 
problem better than those that get easily stuck. 

Many interesting optimization problems are 
discrete in nature, meaning that the variables 
involved are represented by integers, typically 
taking up only two values, say $0$ and $1$. 
A large body of work has thus exploited the 
idea of using Ising-type physical systems 
to effectively map integer variables onto 
binary spins and then use Monte Carlo methods or 
alike to simulate the physical dynamics of these 
analogue Ising machines to find their ground state. 
These techniques have been studied extensively 
in the literature (see, for example, references ~\cite{mezardmontanari,moore, krzakala,Mohseni}).

In this article, we present a new energy-minimization framework to solve combinatorial optimization problems 
based on the physical dynamics of real macrospins 
described by the Landau-Lifshitz-Gilbert (LLG) 
equations. 
The idea is to use 3-dimensional rotors as 
the analogue physical dynamical system, 
rather than up-down Ising spins. The rationale behind our approach is the physical gambit that trajectories away from the z-axis, in particular, those occurring halfway between the north and south poles of the unit sphere will help the system to escape from local solutions, thus facilitating the discovery of the true ground-state. 

We explain our method in two steps: first, we introduce 
the stochastic LLG dynamical system and discuss its numerical 
implementation; second, we explain how to apply 
it to a specific QUBO problem, called the Sherrington–Kirkpatrick 
model in statistical physics~\cite{skmodel, arora05}, which can be 
exactly solved in the thermodynamic limit~\cite{Parisi_1980}, 
and show that our method extrapolates to this exact 
value while performing systematically better than 
discrete methods based on Glauber dynamics of Ising 
spins~\cite{glauber}. 

\begin{figure}[t]
\includegraphics[width=0.5\textwidth]{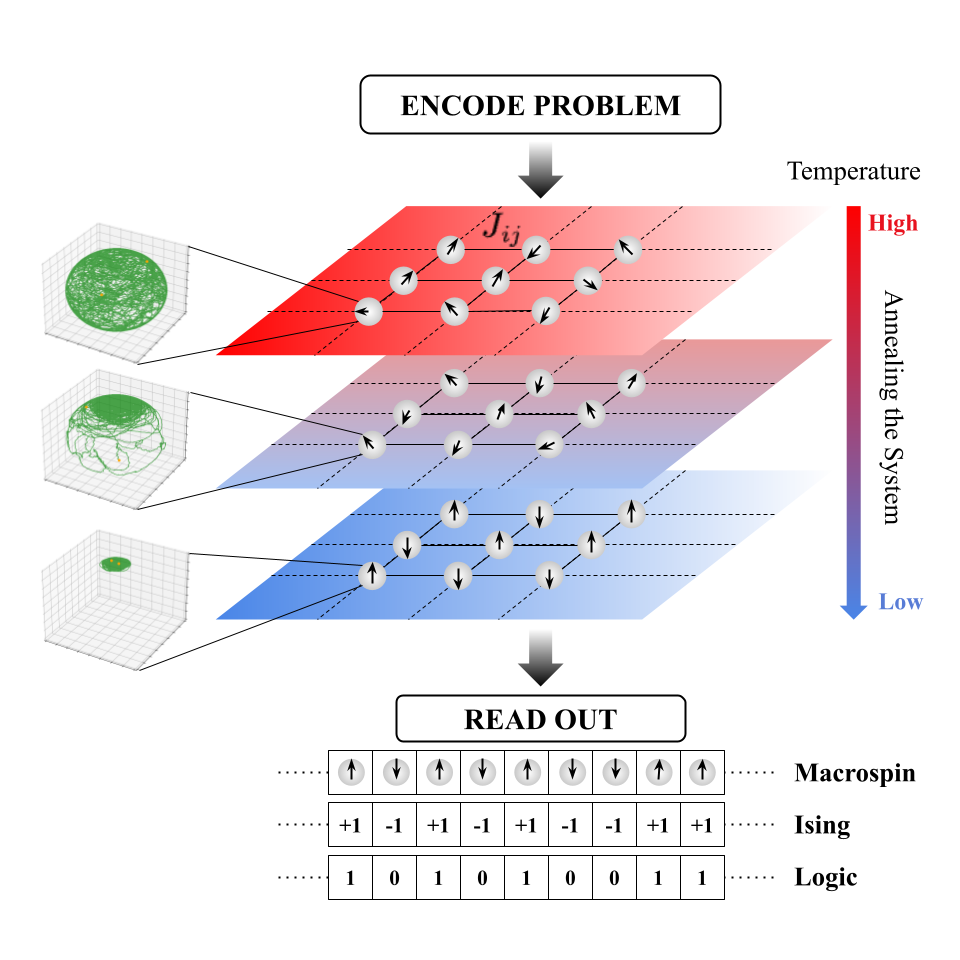}
\caption{{\bf Schematic description of our method}. 
 First, we encode the solution of the optimization 
 problem into the ground state of the spin system's Hamiltonian. 
 The parameters $J_{ij}$ model the interactions between 
 pair of macrospins. Here we only show couplings between 
 neighbouring spins for illustration purposes. In general, 
 the connections are not restricted to the nearest neighbours.  
 Anisotropy is set along the $z$ axis of each spin via a magnetic anisotropy field $H_{Ai}$ favoring the north-south direction. After initializing the macrospins in random directions 
 on the unit sphere, the system is cooled from high to 
 low temperature. At high temperatures, each 
 macrospin experiences large thermal fluctuations with trajectories 
densely covering the unit sphere. As the system cools to lower 
 temperatures, the fluctuations are reduced and the spins 
 trajectories tend to confine into small regions around the 
 poles of the unit sphere. The dynamical 
 trajectories of a single macrospin during the annealing process 
 is depicted in green on the left-hand side of the illustration. 
 Eventually, the system reaches the end of the annealing schedule 
 and spins stabilize around the north and south poles. 
 The spins' final states are binarized based on their up or down 
 direction to get a direct readout of the solution to the original problem. 
}
\label{fig:explanation}
\end{figure}

\begin{figure}
 \centering
 \includegraphics[width=0.5\textwidth]{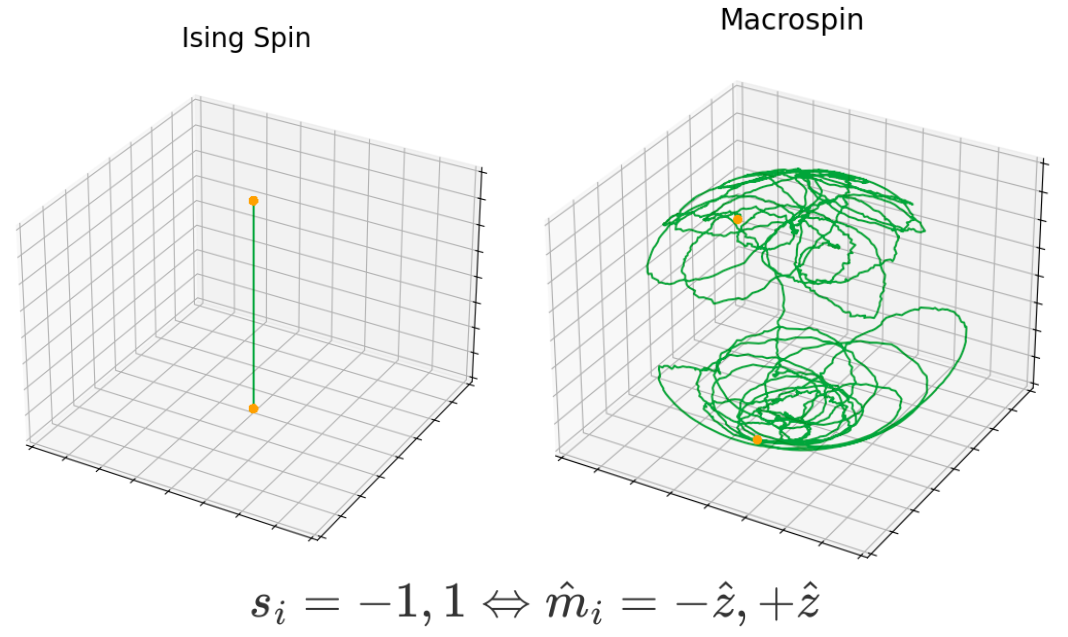}
 \caption{Illustration of Ising spin and macrospin. Ising spins can 
 only take two values, $s_i=\pm 1$. Macrospins, on the other hand, have extra degrees of freedom when rotating on the unit sphere.} 
 \label{fig:ising and macro}
\end{figure}

\emph{Stochastic-LLG Dynamics.} 
We consider a system consisting of $n$ single-domain 
ferromagnets, and apply the macrospin approximation 
to each domain. The magnetization of the spins in the 
$i^\mathrm{th}$ domain are averaged to a single giant 
magnetic moment $\Vec{m}_i = m_i\hat{m}_i$, with magnitude 
$m_i>0$ and direction $\hat{m}_i$, as illustrated 
in Fig.~\ref{fig:explanation}. 
The micromagnetic energy $E$ of our $n$ macrospins system 
has three terms: the energy from the exchange interaction 
between each moments $E_\text{ex}$, the energy from the 
magnetic anisotropy determined by the crystallographic 
structure of the domain, $E_\text{anis}$, and the Zeeman 
energy from an external magnetic field applied to the system,  
$E_\text{Z}$. Summarizing:
\begin{align}
    E = E_\text{ex} + E_\text{anis} + E_\text{Z}.
    \label{eq:micromag energy}
\end{align}
The expression for each term is given by:
\begin{align}
    E_\text{ex} &= -\mu_0 \sum_{\langle i,j \rangle}^{n} 2J_{ij}H_\text{ex} \frac{\Vec{m}_i \cdot \Vec{m}_j}{m_i+m_j},\label{eq:E_ex} \\
    E_\text{anis} &= -\frac{1}{2}\mu_0 \sum_{i=1}^n H_{Ai} m_i (\Vec{e}_{hi} \cdot \hat{m}_i)^2, \label{eq:E_anis} \\
    E_\text{Z} &=-\mu_0 \sum_{i=1}^n \Vec{m}_i\cdot\Vec{H}_{\text{ext},i}, 
    \label{eq:E_ext}
\end{align}
where $\mu_0$ is the vacuum permeability, $H_{Ai}$ 
characterizes the magnetic anisotropy of moment $i$, 
and $\Vec{e}_{hi}$ is the unit vector characterizing 
the preferred crystallographic direction of moment $i$ 
(which later we take to be along the z-axis, see 
Fig.~\ref{fig:ising and macro}). 
The summation over $\langle i,j \rangle$ is over 
pairs of connected spins. 
Vector $\Vec{H}_{\text{ext},i}$ is an external local 
magnetic field acting on the $i^\mathrm{th}$ magnetic moment. $H_\text{ex}$ has units of field and it characterizes the general scale of the exchange coupling strength. Coupling terms $J_{ij}$ are dimensionless and characterize the exchange interaction between magnetic moments. 

Each magnetic moment $\Vec{m}_i$ evolves according to the 
Landau-Lifshitz-Gilbert (LLG) equation~\cite{landau}:
\begin{align}
    \frac{d\Vec{m}_i}{dt} = \gamma_i' \Vec{m}_i \times \Vec{H}_{\text{eff},i} - \frac{\alpha_i}{m_i} \Vec{m}_i \times  \frac{d\Vec{m}_i}{dt},
    \label{eq:LLG}
\end{align}
where $\gamma_i' = \mu_0 \gamma_i$, with $\gamma_i$ the gyromagnetic 
ratio of the $i^\mathrm{th}$ moment, and $\alpha_i$ is the damping constant. 
Vector $\Vec{H}_{\text{eff},i}$ is the effective field experienced 
by the $i^\mathrm{th}$ magnetic moment, given by 
\begin{align}
    \Vec{H}_{\text{eff},i} = -\frac{1}{\mu_0}\frac{\partial E}{\partial \Vec{m}_i}.
    \label{eq:Heff}
\end{align}

To incorporate thermal fluctuations into the dynamics we 
add a thermal field $\Vec{H}_{\text{th},i}$ to $\Vec{H}_{\text{eff},i}$ 
by the substitution
\begin{align}
    \Vec{H}_{\text{eff},i} \rightarrow \Vec{H}_{\text{eff},i} + \Vec{H}_{\text{th},i}.
\end{align}
The thermal fields are delta-correlated zero mean random variables
in time, with a correlation given by~\cite{brown_thermal_1963,GarciaLazaro98}:
\begin{gather}
    \langle \Vec{H}_{\text{th},i,\mu}(t)\Vec{H}_{\text{th},i,\nu}(t')\rangle = C_i \delta_{\mu\nu} \delta (t-t'), \\
    C_i = \frac{2\alpha_i k_B T}{\mu_0^2 m_i \gamma_i(1+\alpha_i^2)},
\label{eq:Brown_space_time}
\end{gather}
where $\mu,\nu=x,y,z$, $k_B$ is the Boltzmann constant, and $T$ 
is the temperature. Integrating Eq.~\eqref{eq:Brown_space_time} over 
a small time interval $\Delta t$, we obtain the standard deviation 
of $\Vec{H}_{\text{th},i}$ as
\begin{align}
    \sigma_i = \frac{1}{\mu_0} \sqrt{\frac{2\alpha_i k_B T}{m_i \gamma_i (1+\alpha_i^2) \Delta t}}.
    \label{eq:H_th_sigma}
\end{align}
In practice, for each component of $\Vec{H}_{\text{th},i}$, we 
draw a value from a Gaussian distribution with mean $\mu = 0$ and 
standard deviation given by $\sigma_i$ in Eq.~\ref{eq:H_th_sigma}. 
Finally, to integrate the stochastic differential equation we apply Heun's integration scheme~\cite{ament_solving_2016}.

In our simulation, we assume that each macrospin $i$ has 
the same crystallographic structure. All parameters 
related to the material properties are the same across 
all $n$ domains, therefore the index $i$ of the following 
quantities are omitted: $H_{Ai} = H_A$,  $\alpha_i = \alpha$, 
$\gamma_i = \gamma$, $\gamma_i' = \gamma'$, and the 
crystallographic direction for each spin is set along the 
z-axis, $\Vec{e}_{hi} = \hat{z}$. We also assume each magnetic 
moment has the same magnitude, $m_i=m$ for all $n$ macrospins. 
This simplification models an array of perfectly identical 
macrospins. The simulation can be further simplified to only 
consider the evolution of the unit magnetic moment $\hat{m}_i$. 
Given the assumptions above, we can think of our model as a system 
of $n$ identical dynamical bits. At the end of the simulation, 
a projection on the z-axis, given by $s_i\equiv{\rm sign}(\hat{m}_{i,z})$, 
maps the macrospin onto a binary Ising-like spin $s_i=\pm1$, 
as illustrated in Fig.~\ref{fig:ising and macro}. 

The values of the parameters used in our simulation 
are listed in Table~\ref{tab:params}. The values we chose here 
are within those of the fluctuating, the so-called free layer, 
of magnetic tunnel junction nanopillars~\cite{Kent2015}.

\begin{table}[h]
\centering
\begin{tabular}{|c|c|c|}
\hline
Damping Constant & $\alpha$ & 0.01 \\
\hline
Magnetic Moment & m & $8\times10^{-19}$J/T \\
\hline
Ambient Temperature & $T_\text{amb}$ & 300K \\
\hline
Exchange Field & $H_\text{ex}$ &  $8.24\times10^3$ A/m\\
\hline
Anisotropy Field & $H_A$ &  $8.24\times10^3$ A/m \\
\hline
Simulation Time Step & $\Delta t$ & $0.01$ ns \\
\hline
Iterations at Each Temperature &   & $2 \times 10^6$ \\
\hline
\end{tabular}
\caption{Values of the parameters chosen for simulation.}
\label{tab:params}
\end{table}
With these parameters, we implement an annealing 
schedule to mimic the physical cooling of the system 
from ambient temperature down to the low-temperature regime. 
Ideally, the annealing schedule should be performed slowly 
from $T_\text{amb}=300$ to $0$~K. 
In practice, we choose an annealing schedule with $n_T=30$ 
intermediate temperatures $T_1>T_2>\cdots>T_{n_T}$ drawn from 
a distribution $P(T)\propto 1/T^{\delta}$, with $\delta=1.1$, 
in order to have more values in the low temperature regime. 
Having explained how to simulate the general 
stochastic-LLG dynamical system, we move next 
to discuss its application to the SK model.

\begin{figure}[h]
\includegraphics[width=0.7\linewidth]{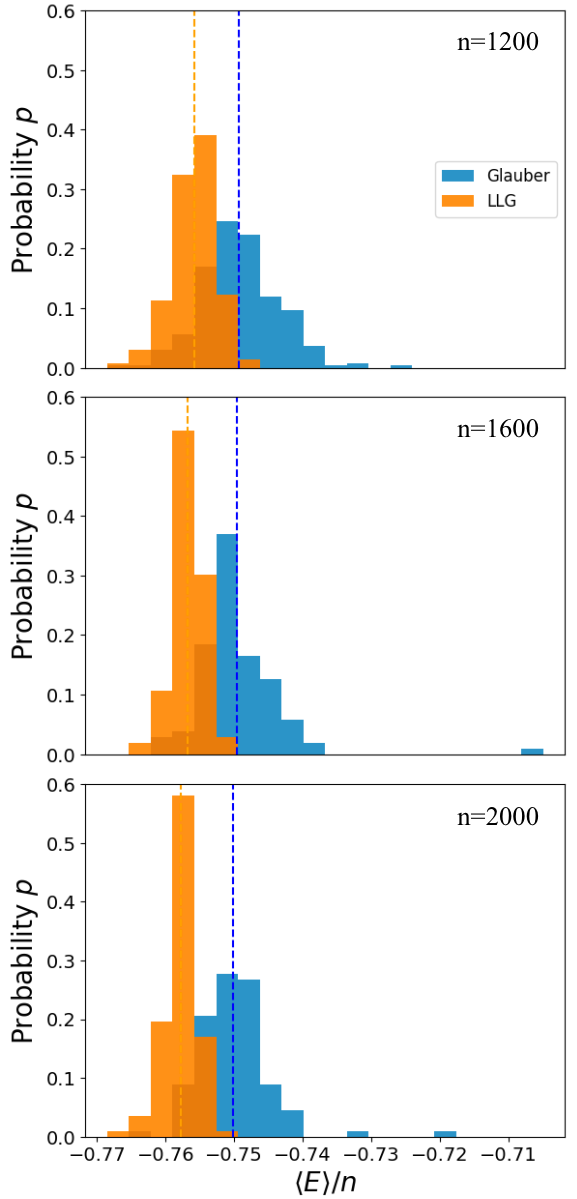}
\caption{Histograms of the final energies $\langle E \rangle/n$ 
obtained from LLG dynamics (orange bars) and Glauber dynamics 
(blue bars) for $n=1200, 1600$, and $ 2000$. 
The colored vertical dashed lines represent the 
average values of the corresponding distributions. 
To make the histograms we used 300 samples for $n=1200$, $103$ for $n=1600$, and $112$ for $n=2000$.
}
\label{fig:hisograms}
\end{figure}

\emph{Results.}
Here we show the simulation results for the SK model 
using the LLG dynamics described above. We also run a Glauber 
dynamics for comparison~\cite{glauber}. The Hamiltonian of the 
SK model for a $n$-spin system is given by:
\begin{align}
    H_{SK}(s) = -\sum_{\langle i,j \rangle}^n \frac{J_{ij}}{\sqrt{n}}\ s_i s_j,
    \label{eq:H_SK}
\end{align}
where $s_i=\pm1$ are Ising spins, $J_{ij}$ is the exchange interaction 
between the $i^\mathrm{th}$ and $j^\mathrm{th}$ spins. Couplings $J_{ij}$ are independent and identically distributed random 
variables drawn from a Gaussian distribution with zero mean and variance 
$\sigma^2 = 1$. 
The summation only counts a pair of $\langle i,j \rangle$ 
once. 
The ground state energy $E_\text{min}\equiv\min_{s} H_{SK}(s)$ 
has been calculated analytically in the limit of infinite system size 
$n \rightarrow \infty$~\cite{Parisi_1980} and is given by 
\begin{equation}
\lim_{n\to\infty} \frac{E_\text{min}}{n} = -0.76321...\ . 
\label{eq:Parisi}
\end{equation}
We can map the Hamiltonian $H_{SK}$ onto our LLG model by directly mapping 
the $J_{ij}$ terms in Eq.~(\ref{eq:H_SK}) to the $J_{ij}$ terms in Eq.~\eqref{eq:E_ex}. The $\Vec{H}_{\text{ext},i}$ term in Eq.~(\ref{eq:E_ext}) is set to zero since there is no external field applied in the SK model. 
With the anisotropy term favoring each spin to align along 
the z-axis, $\Vec{e}_{hi} = \hat{z}$, the ground state of the SK model is mapped onto 
the ground state of the LLG system once we binarize the 
macrospins using their z-projections as illustrated in 
Fig.~\ref{fig:ising and macro}. 
Thus, the analogue LLG Hamiltonian reads
\begin{align}
H_{SK}^{LLG} &= -2\mu_0 H_\text{ex}\sum_{\langle i,j \rangle}^{n} J_{ij}\frac{\Vec{m}_i \cdot \Vec{m}_j}{m_i+m_j} \\
    &-\frac{1}{2}\mu_0 \sum_i^n H_{Ai} m_i (\hat{z} \cdot \hat{m}_i)^2\ . \nonumber
    \label{eq:H_sk of LLG}
\end{align}
We solve the SK model for several system sizes, ranging 
from $n=40$ to $2000$, using both Glauber and LLG dynamics 
and record the value of the final energy $E$ obtained with 
both methods. 
As shown in Fig.~\ref{fig:hisograms}, our LLG 
method performs better, {\em i.e.}, lower energies 
are consistently found for any value of $n$. 

\begin{figure}[h!]
 \centering
 \includegraphics[width=0.5\textwidth]{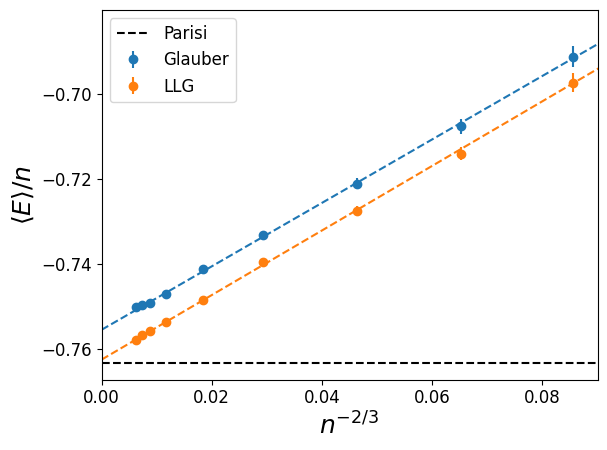}
 \caption{Average ground state energy per spin 
 $\langle E \rangle/n$ as a function of $n^{-2/3}$~\cite{Slanina} 
 for Glauber (blue line) and LLG (orange line) 
dynamics. The horizontal dashed line represents the asymptotic $n\to\infty$ exact ground state energy given by Eq.~\eqref{eq:Parisi}. 
Data points from the LLG dynamics plausibly extrapolate 
to the true optimal value. Error bars indicate the 
standard error of the mean energy per spin.} 
 \label{fig:llg vs glauber}
\end{figure}

In Fig.~\ref{fig:llg vs glauber} we plot the average 
ground state energy per spin, $\langle E \rangle /n$, 
obtained from Glauber and LLG as a function of 
$n^{-2/3}$~\cite{Slanina}.
We observe that the LLG dynamics results extrapolate to 
$\langle E \rangle /n = -0.762(1)$ as $n\to\infty$, 
which is very close to the exact optimal value 
Eq.~\eqref{eq:Parisi}, while the Glauber dynamics 
results intercept the $y$ axis at a higher sub-optimal 
asymptotic value $\langle E \rangle /n = -0.755(2)$.  

Recently it is has been shown, assuming a widely believed conjecture about the SK model, that there is a polynomial time algorithm for finding (with high probability) a string whose energy is between the lowest energy and $(1-\epsilon)$ of the lowest energy of the problem~\cite{Montanari19}. Our results suggest we can achieve similar performance.  

As a point of reference, if the algorithm would get stuck 
at the critical point of the SK model (at $T=1$) the typical energy per spin would be $\langle E \rangle /n=-0.5$. However, even greedy spin flip dynamics, in which one only accepts flips if they lower the energy, results in a energy of around $\langle E \rangle /n \approx -0.71$ and this algorithm terminates in a time of $O(n)$. In contrast, spectral relaxation in which one binarizes the lowest energy eigenvector of $J_{ij}$ only achieves an energy of $\langle E \rangle /n \approx -0.637$ while being computationally more costly~\cite{aizenman87}. Finally, quantum methods have been proposed to solve the SK problem~\cite{googleSK,Farhi2022quantumapproximate}. In particular, rigorous analysis of the Quantum Approximate Optimization Algorithm (QAOA) in Ref.~\cite{Farhi2022quantumapproximate} shows that at $p=11$ (with $p$ being the number of layers in the quantum circuit) one can outperform the spectral relaxation result. Although the limit of $p\rightarrow \infty$ is not established, the results in \cite{Farhi2022quantumapproximate} indicate convergence to the Parisi value like $1/p^a$ where $a=O(1)$. However, the cost of evaluating the circuit parameters is of $O(16^p)$ such that one incurs an exponential cost in $1/\epsilon$ in order to get $\epsilon$ close to the ground state. While this cost might be brought down, it would currently be prohibitively expensive to reach the LLG results as $p$ would have to be about $p\sim 40$. Interestingly, the LLG dynamics is simply the semi-classical approximation of the real quantum dynamics of these spins, suggesting that we still have much to learn about these quantum algorithms.  

As in the case of quantum annealing, our method is important because it offers the possibility of being 
realized experimentally in the lab by means of arrays of 
magnetic tunnel junction devices, whose physical 
dynamics is governed precisely by the LLG equations. These devices operate at nanosecond timescales and consume little power. 
Interesting future research directions would be to 
understand the statistical limits of the robustness 
of our method across different choices of the system's 
parameters, as well as the inclusion of 3-spin interactions 
to solve 3-SAT problems, which are of central importance 
in the vast context that stretches from artificial intelligence, 
to cryptography, and operations research.

\medskip

\emph{Acknowledgements.}
We thank Jonathan Z. Sun and Daniel L. Stein for helpful discussions.
This work was supported in part through the NYU IT High Performance 
Computing resources, services, and staff expertise. 
This research was supported by the Office of Naval Research (ONR) under 
Award No. N00014-23-1-2771. 
The  Flatiron  Institute  is  a  division  of  the  
Simons Foundation. We acknowledge support from Air 
Force Office of Scientific Research(AFOSR): Grant FA9550-21-1-0236.

\begin{comment}
\emph{Additional information.}

{\bf  Code availability}
The source code to solve the LLG equations is 
available upon request. 

\medskip

{\bf Acknowledgements}
We thank Jonathan Z. Sun and Daniel L. Stein for helpful discussions.
This work was supported in part through the NYU IT High Performance 
Computing resources, services, and staff expertise. 
This research was supported by the Office of Naval Research (ONR) under 
Award No. N00014-23-1-2771. 
The  Flatiron  Institute  is  a  division  of  the  
Simons Foundation. We acknowledge support from Air 
Force Office of Scientific Research(AFOSR): Grant FA9550-21-1-0236. 

\medskip

\noindent
{\bf Author contributions}
All authors contributed equally to this work. 

\medskip

\noindent
{\bf  Competing interests} 
All authors declare no competing interests. 

\medskip

\noindent
{\bf Correspondence} should be addressed to F. M. at: fm2452@nyu.edu
\end{comment}

\bibliography{mybib}

\end{document}